\documentclass[]{spie}  

\usepackage{amsmath,amsfonts,amssymb}
\usepackage{graphicx}
\usepackage{booktabs}
\usepackage{enumitem}
\usepackage{float}          
\usepackage[ruled,vlined]{algorithm2e}
\usepackage[colorlinks=true, allcolors=blue]{hyperref}
\usepackage{subcaption}  

\title{From Narrow to Wide: Autoencoding Transformers for Ultrasound Bandwidth Recovery}

\author[a]{Sepideh K. Gharamaleki}
\author[a]{Hassan Rivaz}
\author[b, c]{Brandon Helfield}

\affil[a]{Concordia University, Department of Electrical and Computer Engineering, Montreal, QC, H3G 2W1, Canada}
\affil[b]{Concordia University, Department of Physics, Montreal, QC, H4B 1R6, Canada}
\affil[c]{Concordia University, Department of Biology, Montreal, QC, H4B 1R6, Canada}

\begin{document} 
\maketitle

\begin{abstract}
Conventional pulse-echo ultrasound suffers when low-cost probes deliver only narrow fractional bandwidths, elongating pulses and erasing high-frequency detail.  
We address this limitation by learning a data-driven mapping from band-limited to broadband spectrogram of \emph{radio-frequency} (RF) lines. To this end, a variation of Tiny Vision Transform (ViT) auto-encoder is trained on simulation data using a curriculum-weighted loss. On heterogeneous speckle–cyst phantoms, the network reduces image-domain MSE by \textbf{90\%}, boosts PSNR by \textbf{+6.7 dB}, and raises SSIM to \textbf{0.965} compared with the narrow-band input.  
It also sharpens point-target rows in a \emph{completely unseen} resolution phantom, demonstrating strong out-of-distribution generalisation without sacrificing frame rate or phase information. These results indicate that a purely software upgrade can endow installed narrow-band probes with broadband-like performance, potentially widening access to high-resolution ultrasound in resource-constrained settings.
\end{abstract}


\section{INTRODUCTION}
\label{sec:intro}

Ultrasound (US) remains the workhorse of diagnostic imaging because it is real‑time, portable, and relatively inexpensive.  Yet the spatial resolution and contrast of any US system are fundamentally limited by the fractional bandwidth of its transducer: the broader the bandwidth, the shorter the emitted pulse, and the sharper the resulting axial resolution and speckle statistics. Conversely, probes with narrow bandwidths inherently emit longer pulses, which blur small structures and can mask clinically relevant details, especially in applications such as microvascular imaging \cite{Gharamaleki2025_DeformableDETR_ULM}. On the other hand, clinical scenarios that demand greater penetration depth, for example abdominal imaging, require a lower transmit frequency ($\sim$1–3 MHz) to curb frequency-dependent attenuation and preserve echo strength \cite{Kremkau2019}. Unfortunately, high‑bandwidth piezo‑composite arrays dramatically raise manufacturing cost and system complexity, which restricts their deployment in resource‑constrained settings and older scanners.

Early research attempted to compensate for bandwidth loss in hardware by stacking multiple piezo layers \cite{He2023_SharedDualFreqIVUS}, or switching to more modern technology; although these approaches widen the spectrum, they require new probes and front‑end electronics.  Signal‑processing alternatives such as frequency compounding \cite{Chang2010_FreqCompounding} and resolution‑enhancement compression \cite{Lei2024_LFUQI_CNN} improve speckle and axial resolution by synthesising multiple sub‑band images, but they sacrifice frame rate and cannot recreate truly missing high‑frequency content.

The emergence of deep learning has reframed bandwidth compensation as a supervised translation task: a network learns a mapping from narrow‑ to wide‑band images (or radio‑frequency data) using paired examples.  Awasthi \emph{et al.} demonstrated up to a 3.9 dB PSNR gain when transforming 20\%‑bandwidth B‑mode images to their 100\%‑bandwidth counterparts with U‑Net and REDNet \cite{Awasthi2023_BandwidthImprovement}. Robins \emph{et al.} further showed that CNN‑based extrapolation of missing low‑frequency components in full‑waveform inversion improves breast US tomography reconstructions \cite{Robins2021_DLFWI_USBreast}. Despite these advances, existing studies either operate on beam‑formed B‑mode images, where phase information is irreversibly lost, or target transmission tomography systems that differ from conventional pulse‑echo imaging.

To our knowledge, no prior work has tackled \emph{spectral super‑resolution} for conventional pulse‑echo US: i.e.\ reconstructing the broadband radio-frequency lines that a wide-bandwidth probe would record, using only the beamformed echoes captured by a narrow-band transducer at the same spatial position. Solving this problem would instantly upgrade the effective resolution of millions of installed probes without hardware changes, unlock finer depiction of micro‑structures, and harmonise data across heterogeneous clinical devices.

We formulate bandwidth extension as a learning problem on \emph{paired} simulated RF data.  A broadband ground‑truth dataset is generated with \textsc{Field II}   and then numerically band‑limited to emulate a low‑cost probe.  Inspired by MAE-AST \cite{Baade2022_MAEAST}, which recovers spectograms of audio datasets, we train a ViT-based encoder-decoder architecture to recover the missing spectral content directly in the RF domain.  Extensive experiments on speckle and resolution phantoms, show that our method boosts PSNR, SSIM, CNR and SNR while recovering the lost frequencies.

\section{METHODS AND MATERIALS}
\label{sec:methods}

\subsection{Data Acquisition}

All RF data were generated with the \textsc{\textsc{Field II}  }  
programming interface by scanning a set of 1800 independent point‑scatterer phantoms similar to \cite{Sharifzadeh2024_MitigatingAberrationNoise}. Every phantom file contains \(3\times10^{5}\) scatterers with randomised positions and Rayleigh‑distributed amplitudes. By deriving the masks from real-world images, the training data remains agnostic to any particular anatomical or structural pattern, preventing the network from developing bias toward specific shapes. 
For the simulations, we used a linear array with 192 elements with a center frequency of \(5.208~\text{MHz}\) and sampling frequency \(f_{s}=20.832~\text{MHz}\), and tapered by a Hanning window.

\subsubsection{Low vs. High Bandwidth Simulation}
\textsc{Field II}   allows two independent levers for shaping the transmitted spectrum:
(i)~the piezo‑electric impulse response of the probe, and 
(ii)~the electrical excitation waveform.  
Instead of defining multiple probe models with different –6 dB fractional
bandwidths, we modelled a \textbf{single idealised transducer} whose –6 dB
pass‑band spans twice the nominal baseline bandwidth (\text{2BW}).
Exciting a broadband transducer with spectrally restricted pulses lets us mimic a narrow-band probe while keeping its aperture geometry, beam directivity, and focal profile unchanged.

Let \(H_{\text{probe}}(f)\) denote the probe’s transfer function and
\(E(f)\) the Fourier transform of the excitation voltage.  
The transmitted pressure spectrum is 
\(P(f)=H_{\text{probe}}(f)\,E(f)\). The probe impulse response was generated with \textsc{Field II}  ’s
\verb|impulse_response| utility using a Hann‑windowed sinusoid centred at
\(f_{c}=5.2\;\text{MHz}\) and a \(-6\;\text{dB}\) fractional bandwidth of
\(2\text{BW}=120\%\) (i.e.\ 60 \% on either side of \(f_{c}\)).
Two Gaussian‑modulated excitations were then defined:

\begin{itemize}[leftmargin=1.5em]
    \item \textbf{Wide‑band excitation (\(2\text{BW}\)).}  
          Standard deviation \(\sigma_{2}\) chosen such that the
          \(-6\;\text{dB}\) points coincide with \(\pm60\%\) of \(f_{c}\).
    \item \textbf{Narrow‑band excitation (\(\text{BW}\)).}  
          Standard deviation \(\sigma_{1}=\sigma_{2}/\sqrt{2}\) to limit the
          spectrum to \(\pm30\%\) of \(f_{c}\).
\end{itemize}

Because \(E_{\text{BW}}(f)=0\) for \(|f|>0.3f_{c}\), the product  
\(H_{2\text{BW}}(f)E_{\text{BW}}(f)\) equals the pressure spectrum that would
be obtained from a true BW‑limited probe excited with \(E_{\text{BW}}(f)\).
This strategy therefore, yields two consistent datasets:

\[
\boxed{\text{Low‑BW simulation: } \;H_{2\text{BW}}\times E_{\text{BW}}}
\quad\text{and}\quad
\boxed{\text{High‑BW simulation: } H_{2\text{BW}}\times E_{2\text{BW}} }.
\]

\subsection{Short‑time Fourier Transform And Spectrogram Construction}
Inspired by audio-spectrogram inpainting methods that infer masked frequency bands (e.g., \cite{Gong2022_SSAST}), we aim to reconstruct the high-bandwidth components of ultrasound signals from their bandwidth-limited equivalents. Specifically, every RF line from the simulated data is subjected to a 2-D FFT, after which the spectrogram’s magnitude and phase are stored as separate tensors that feed the reconstruction model.

To calculate the spectograms of the RF data lines, Single--channel RF traces of length $L=1536$ were analyzed using Short Time Fourier Transform (STFT) which divides the signal into short, overlapping windows and assumes stationarity within each window. The parameters of STFT, were chosen to satisfy the spectral‑resolution requirements of the system while remaining computationally efficient. Table~\ref{tab:stft-params} summarizes these configurations.

\begin{table}[h!]
    \centering
    \caption{STFT parameter settings.}
    \label{tab:stft-params}
    \begin{tabular}{c|c}
        \hline
        Parameter & Value \\
        \hline
        \(N_{\mathrm{FFT}}\) & 512 \\
        Window length \(w\) & 64 \\
        Hop size \(h\) & 32 \\
        Window type & Hamming \\
        \verb|center| & True (pad \(w/2\) on both sides) \\
        \verb|onesided| & True (keep only nonnegative frequencies) \\
        \hline
    \end{tabular}
\end{table}

\subsection{Network Architecture}

Let \(x \in \mathbb{C}^{T \times F}\) denote the complex STFT of a single RF line, where \(T\) and \(F\) are the time- and frequency-axis lengths, respectively. We decompose \(x\) into magnitude and phase channels,
\begin{equation}
  X \;=\;
  \bigl[\,|x|,\;\angle x\bigr] \in \mathbb{R}^{2 \times T \times F},
\end{equation}
and apply channel-wise \emph{batch normalisation}. The normalised spectrogram is then partitioned into non-overlapping patches using a 2-D \texttt{Unfold} operation with kernel size
\(k_t \times k_f = 8 \times 8\) and stride \(s_t = s_f = 8\).
Each patch \(\mathbf{p} \in \mathbb{R}^{2k_tk_f}\) is linearly projected to a token
\begin{equation}
  \mathbf{z}_p \;=\; \mathbf{W}_{\text{proj}} \, \mathbf{p}
  \;\in\; \mathbb{R}^{D}, 
  \quad D = 768.
\end{equation}

\begin{figure} [ht]
    \begin{center}
    \includegraphics[height=9cm]{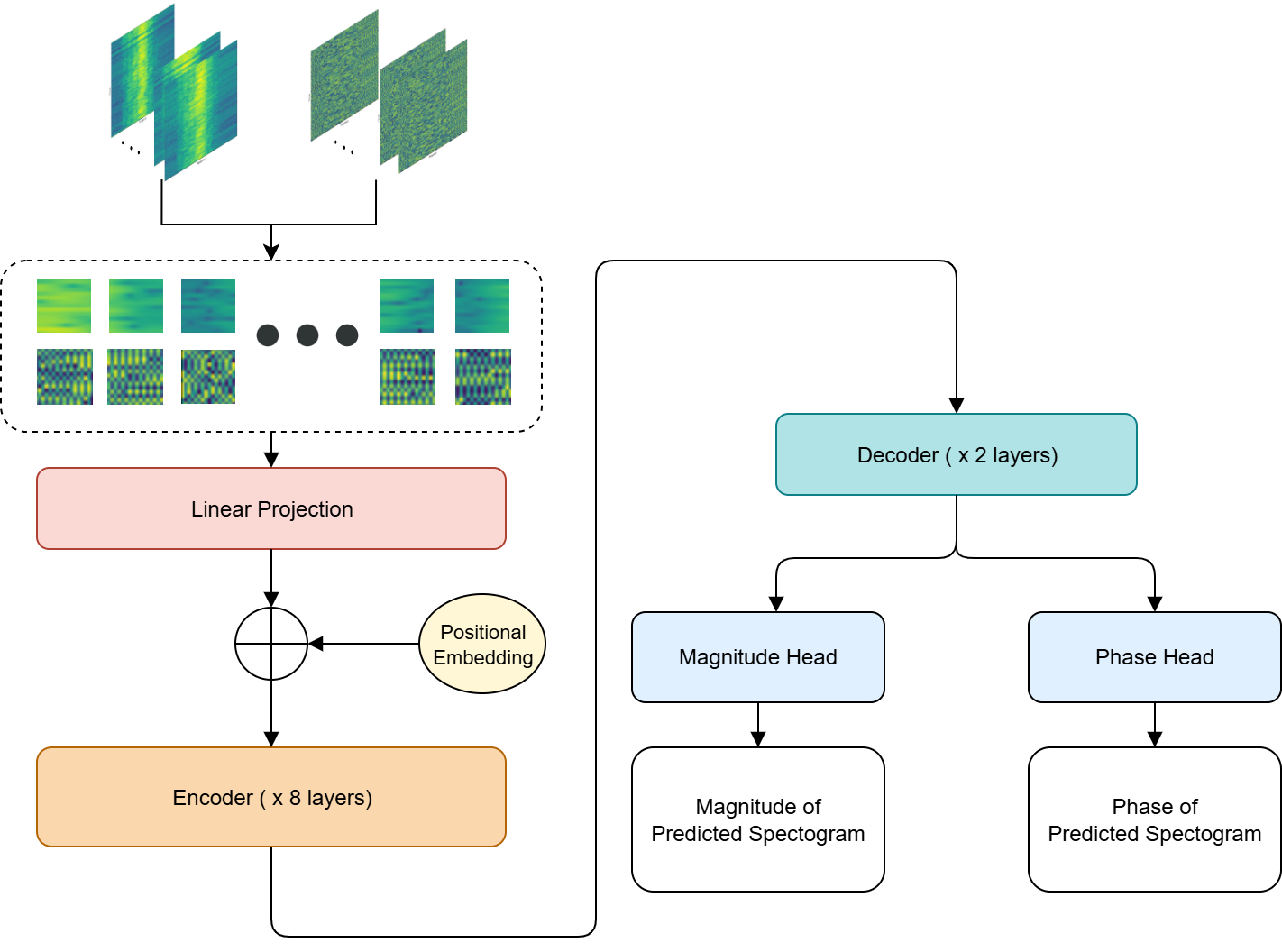}
    
    \end{center}
    \caption{Network architecture visualizing a Tiny ViT-based encoder-decoder structure that reconstructs the phase and magnitude of the higher-bandwidth spectrograms.}
    \label{fig:network_architecture} 

\end{figure}

This token, along with sinusoidal positional encodings of the patches, is then fed to our backbone (Fig.~\ref{fig:network_architecture}) which follows the low-weight encoder–decoder split of MAE-AST, modified for two-channel inputs with fewer encoder layers.
The decoder output \(\mathbf{Y} \in \mathbb{R}^{N \times D}\)
(\(N\) = number of patches) is then mapped back to the patch domain,
\begin{equation}
  \widehat{\mathbf{P}}
  \;=\;
  \mathbf{W}_{\text{recon}} \, \mathbf{Y}
  \;\in\; \mathbb{R}^{N \times 2k_tk_f}.
\end{equation}
a 2-D \texttt{Fold} inverts patchification and averages overlaps, to create the final reconstructed image.

\subsection{Curriculum--Weighted Multi--Term Loss}
\label{sec:curriculum_loss}

\paragraph{Per--component and coupled objectives.}
Each predicted spectrogram pixel is represented in polar form
$\hat{\mathbf{s}}=(\hat{r},\hat{\theta})$, with ground--truth
$\mathbf{s}=(r,\theta)$.
We minimise three complementary mean--squared errors:
\begin{align}
  \mathcal{L}_{\mathrm{mag}}   &= \operatorname{MSE}\bigl(\hat{r},\,r\bigr), \\[-2pt]
  \mathcal{L}_{\mathrm{phase}} &= \operatorname{MSE}\bigl(\hat{\theta},\,\theta\bigr), \\[-2pt]
  \mathcal{L}_{\mathrm{cplx}}  &= \operatorname{MSE}\bigl(\hat{z},\,z\bigr)
                                   =\bigl\lvert\hat{z}-z\bigr\rvert^{2},
\end{align}
where
$\hat{z}=\hat{r}\,e^{j\hat{\theta}}$ and
$z      =      r\,e^{j\theta}$ are the complex predictions and targets,
respectively.

The composite objective at iteration $t$ is
\begin{equation}
  \mathcal{L}(t)=
    \lambda_{\mathrm{mag}}(t)\,\mathcal{L}_{\mathrm{mag}}
  + \lambda_{\mathrm{phase}}(t)\,\mathcal{L}_{\mathrm{phase}}
  + \lambda_{\mathrm{cplx}}(t)\,\mathcal{L}_{\mathrm{cplx}},
  \qquad
  \sum_{k}\lambda_{k}(t)=1.
  \label{eq:total_loss_final}
\end{equation}
Training begins with \emph{axis--aligned} supervision
($\lambda_{\mathrm{mag}},\lambda_{\mathrm{phase}}\!\approx\!1$) and
progressively shifts emphasis to the coupled loss
$\mathcal{L}_{\mathrm{cplx}}$, which enforces global amplitude--phase
coherence. The weights are updated once per epoch using a
data--driven curriculum as follows:

\begin{algorithm}[H]
\caption{Adaptive weight schedule (executed once per epoch $e$)}
\label{alg:adaptive_weights}
\SetKwInOut{Input}{Input}\SetKwInOut{Output}{Output}
\Input{%
  EMA (Estimated Moving Average) coefficient $\beta=0.9$; \\
  minimum and maximum weights $\lambda_{\min}=0.1$ and $\lambda_{\max}=1$; \\
}
\Output{Updated weights
$\lambda_{\mathrm{mag}}^{(e)},\,
 \lambda_{\mathrm{phase}}^{(e)},\,
 \lambda_{\mathrm{cplx}}^{(e)}$}
\BlankLine
\If{$e=0$}{
  Store baselines
  $\mathcal{B}_{p}\leftarrow\bar{\mathcal{L}}_{p}^{(0)}$, \;
  set $\lambda_{\mathrm{mag}}^{(0)}=\lambda_{\mathrm{phase}}^{(0)}=1$,  $\lambda_{\mathrm{cplx}}^{(0)}=0$.\;
}
\Else{
  \ForEach{$p\in\{\mathrm{mag},\mathrm{phase}\}$}{
    Update EMA:
    $\tilde{\mathcal{L}}_{p}^{(e)}
      \leftarrow
      \beta\,\tilde{\mathcal{L}}_{p}^{(e-1)}
      +(1-\beta)\,\bar{\mathcal{L}}_{p}^{(e)}$\;
    Compute remaining--error ratio:
    $r_{t}^{(e)}\leftarrow
      \min\!\bigl(1,\,
        \tilde{\mathcal{L}}_{p}^{(e)} / \mathcal{B}_{p}\bigr)$\;
    Update auxiliary weight:
    $\lambda_{p}^{(e)}
      \leftarrow
      \max\!\bigl(\lambda_{\min},
                  \lambda_{\max}\times r_{p}^{(e)}\bigr)$\;
  }
  Update complex weight:\;
  $\displaystyle
    \lambda_{\mathrm{cplx}}^{(e)}\leftarrow
      \lambda_{\mathrm{cplx},\max}\Bigl[
        1-\tfrac{1}{2}\bigl(r_{\mathrm{mag}}^{(e)}+
                            r_{\mathrm{phase}}^{(e)}\bigr)
      \Bigr]$\;
  Normalize so that $\sum_{k}\lambda_{k}^{(e)}=1$.\;
}
\end{algorithm}

Because the EMA--smoothed ratios $r_{p}^{(e)}$ are monotonically
non--increasing, $\lambda_{\mathrm{mag}}$ and $\lambda_{\mathrm{phase}}$
decay smoothly towards $\lambda_{\min}$, while
$\lambda_{\mathrm{cplx}}$ rises towards
$\lambda_{\max}$ once the auxiliary objectives plateau.

The schedule realises a \emph{smooth, monotone curriculum} that
automatically transfers capacity from easy, decoupled objectives to the
hard coupled objective, accelerating convergence and improving final
spectrogram fidelity.
We trained the network with the AdamW optimizer with a fixed weight-decay term applied to all trainable parameters to discourage overfitting, and, to further stabilize learning, we clipped the global gradient norm at each backward pass. 
\section{Results}
\label{sec:results}

The network was trained exclusively on \textbf{heterogeneous speckle phantoms} that combine
real-world tissue masks with synthetic hypoechoic cysts.
To evaluate \emph{generalisation}, we report results on a speckle-cyst phantom and a point-target resolution phantom, quite different from the training set.

Table~\ref{tab:image-based-metrics} lists the mean\,\(\pm\)\,SD of three standard
image-quality metrics over the speckle phantoms.
Relative to the \emph{low-bandwidth (Low-BW) input}, the network prediction
lowers the MSE by \(\mathbf{90\%}\),
raises the peak signal-to-noise ratio (PSNR) by \(\mathbf{+6.7\;{\rm dB}}\),
and pushes the structural similarity (SSIM) to \(\mathbf{0.995}\), indicating an
almost perfect perceptual match to the broadband ground truth.

\begin{table}[t]
    
  \centering
  \caption{Quantitative comparison between the low-bandwidth \textit{Input}
           and the \textit{Prediction}.}
  \label{tab:image-based-metrics}
  \begin{subtable}{\linewidth}
    \centering
    \caption*{\textbf{Image-Based Metrics}}
    \begin{tabular}{lcc}
      \toprule
      \textbf{Metric} & \textbf{Input} & \textbf{Pred} \\
      \midrule
      MSE ($\times10^{-6}$) & 151.4 ± 4.334 & \textbf{16.547 ± 1.098} \\
      PSNR (dB)            & 38.20 ± 0.12           & \textbf{44.89 ± 0.66} \\
      SSIM                 & 0.9168 ± 0.0021 & \textbf{0.9548 ± 0.0007} \\
      \bottomrule
      \vspace{5pt}
    \end{tabular}
  \end{subtable}
\end{table}

A visual example of the ground truth, prediction and input for a sample speckle–cyst phantom is provided and we further computed contrast-specific
metrics within manually drawn foreground/background ROIs for this example
(Table~\ref{tab:roi_metrics}).  The network restores
both contrast-to-noise ratio (CNR) and signal-to-noise ratio (SNR) to
ground-truth levels, outperforming the Low-BW input by
\(\,+21\%\) and \(+4\;\text{dB}\), respectively.

\begin{table}[t]
  \centering
  \caption{ROI-based metrics for an example speckle–cyst phantom.}
  \label{tab:roi_metrics}
  \begin{tabular}{lccc}
    \toprule
    \textbf{Metric} & \textbf{Ground Truth} & \textbf{Prediction} & \textbf{Input}\\
    \midrule
    SNR (dB)             & 73.83 & \textbf{74.84} & 70.73\\
    CNR  (\%)           & 1.47  & \textbf{1.41} & 1.16\\
    \bottomrule
  \end{tabular}
\end{table}

\begin{figure*}[t]
  \centering
  \newcommand{\panelwidth}{1.0\linewidth}  
  \includegraphics[width=0.9\linewidth,
                 height=.28\textheight,  
                 keepaspectratio=false]  
                 {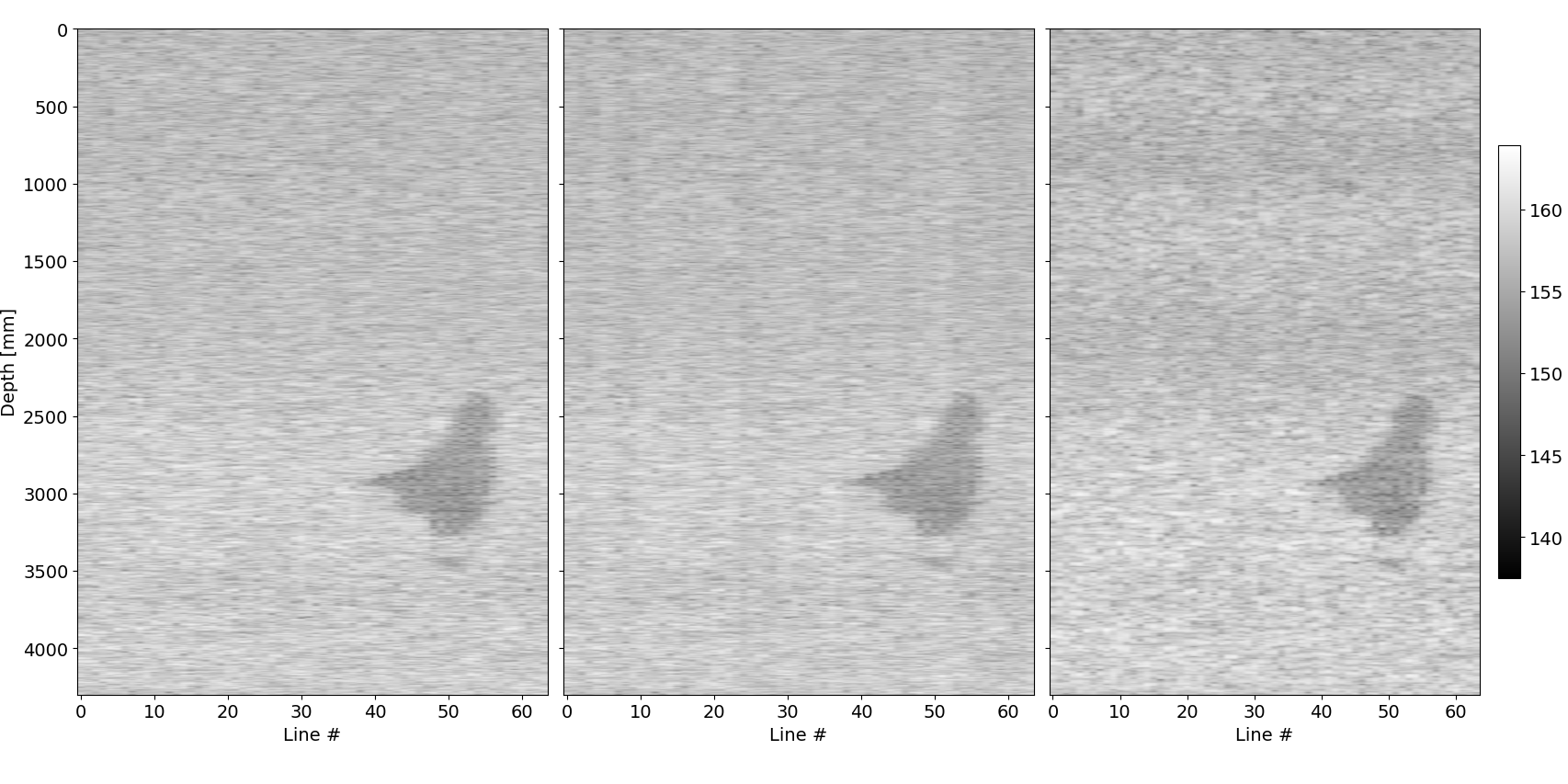}\\[-5pt]

  \begin{tabular*}{0.69\linewidth}{@{\extracolsep{\fill}}ccc}
    \small Ground-truth & \small \textbf{Prediction} & \small Low-BW Input
  \end{tabular*}

  \caption{Performance on the \textbf{speckle–cyst phantom}.
           Columns show (left) broadband ground truth, (centre) network
           prediction, and (right) Low-BW input.
           The proposed method restores speckle granularity and cyst contrast
           that are suppressed in the narrow-band acquisition.}
  \label{fig:speckle}
\end{figure*}

\begin{figure*}[t]
  \centering
  \newcommand{\panelwidth}{1.0\linewidth}  
  \includegraphics[width=0.9\linewidth,
                 height=.28\textheight,  
                 keepaspectratio=false]  
                 {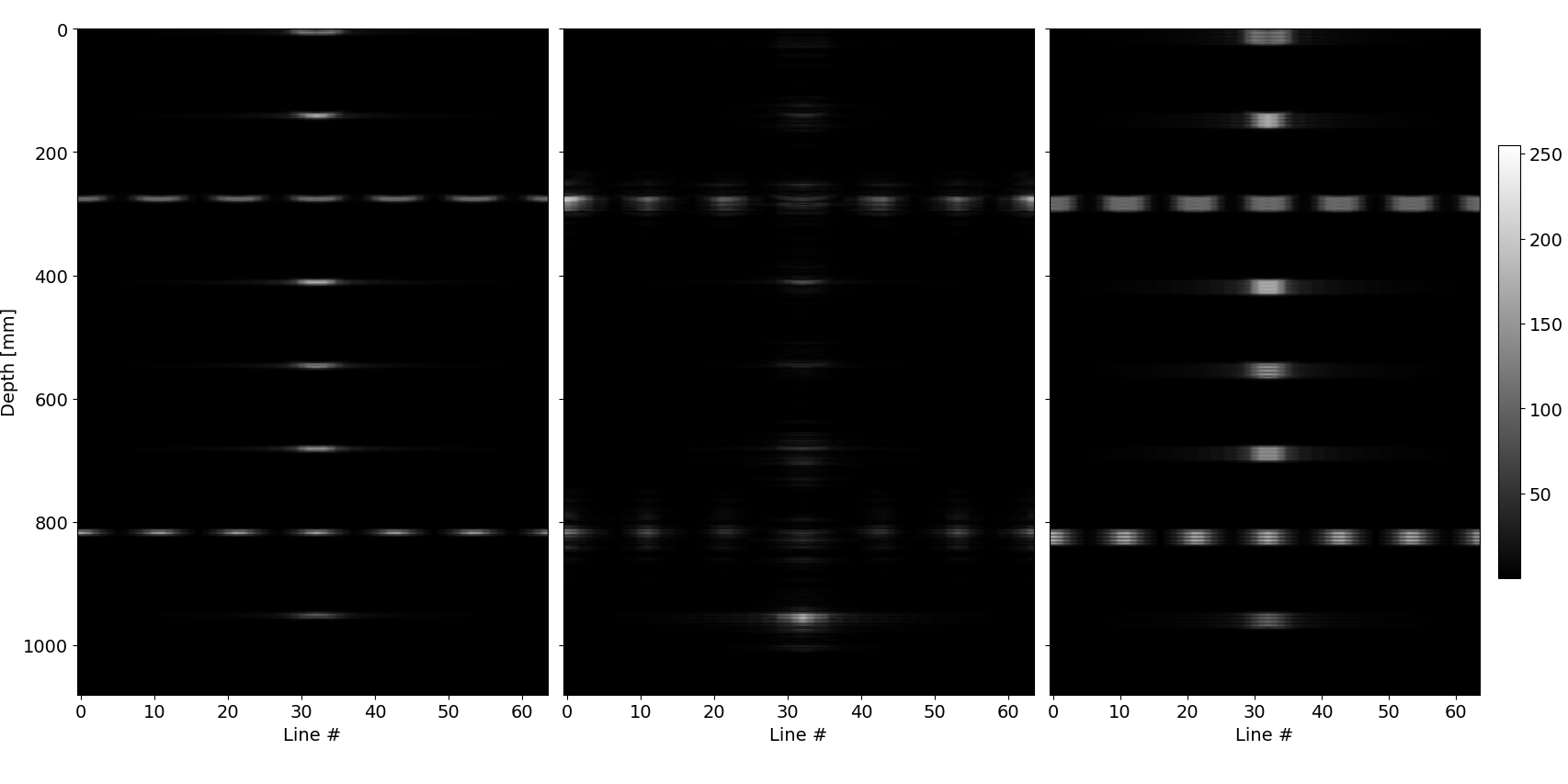}\\[-5pt]

  \begin{tabular*}{0.68\linewidth}{@{\extracolsep{\fill}}ccc}
    \small Ground-truth & \small \textbf{Prediction} & \small Low-BW Input
  \end{tabular*}

  \caption{Performance on the unseen \textbf{point-target resolution phantom}.
           The network sharply reconstructs the sparse scatterer rows that are
           severely blurred in the Low-BW input, evidencing strong
           out-of-distribution generalisation.}
  \label{fig:res}
\end{figure*}

Figure~\ref{fig:speckle} shows B-mode images for the speckle-cyst phantom.
Fine speckle texture and the cyst boundary—blurred in the Low-BW input—are
reconstructed with high fidelity.  
Figure~\ref{fig:res} depicts the unseen resolution phantom: the model
recovers narrow point-spread-function (PSF) rows, reducing MSE by a further
\(\sim3\times\) and boosting PSNR by \(\sim5\;\text{dB}\) over the input.
The full-width at half-maximum (FWHM) of the single targets is visibly sharper,
demonstrating that spectral super-resolution translates into improved spatial
resolution, even on data drawn from a \emph{disjoint} distribution.

\section{Discussion and Conclusion}

We introduced the first transformer-based
auto-encoder for \emph{spectral super-resolution} of pulse-echo ultrasound
RF data.  Trained on synthetically band-limited simulations, the model
recovers missing frequency bands with high fidelity, delivering up to
\(+6.7\,\text{dB}\) PSNR improvement and near-perfect SSIM on heterogeneous
phantoms.  
Although the model was trained solely on speckle-cyst phantoms, it generalises
to a structurally unrelated resolution phantom.  We attribute this behaviour to our physics-consistent approach; By using the RF data spectograms, the network learns the spectral correlations rather than memorising spatial textures. Another advantage of working directly in the RF domain is retaining the phase info while not suffering any frame-rate penalty. These results
suggest that our network can retrofit existing narrow-band probes with
broad-band-like performance through a purely software upgrade, opening a path
toward higher-resolution imaging in resource-constrained clinical settings.

 In the resolution phantom the network occasionally under-estimates the
        amplitude of isolated scatterers, yielding faint points that are harder
        to discern than in the broadband reference. We relate this to the fact that the temporal correlation between the RF lines have been ignored in the network structure. Furthermore, although the plain ViT backbone is parameter-efficient, it must cache
        all \(N\) patch tokens simultaneously, leading to
        \(\mathcal{O}(N^{2})\) attention and memory growth.  For long RF
        traces sampled at \(\ge 40\,\text{MHz}\), this quickly exhausts GPU
        memory.  A sliding-window strategy or a hierarchical transformer
        (e.g.\ Swin or HRFormer) could alleviate this bottleneck without
        sacrificing the receptive field.

        As a future experiment, we will evaluate the results of the network on tissue-mimicking phantoms acquired
        with \emph{two clinically distinct probes}—one broadband and one
        narrow-band—to quantify performance under real acoustic noise,
        cable dispersion, and electronic artefacts.

\bibliography{report} 
\bibliographystyle{spiebib} 

\end{document}